\documentclass[journal]{IEEEtran}

\usepackage{amsmath,amssymb,amsfonts}
\usepackage{graphicx}
\usepackage{cite}
\usepackage{bm}
\graphicspath{{./}}

\usepackage{xcolor}

\newcommand{\E}{\mathbb{E}}
\newcommand{\Hcal}{\mathcal{H}}
\newtheorem{lemma}{Lemma}

\begin{document}

\title{Service-Area Robust NOMA for Pinching \\ Antennas Systems}

\author{Zoran Hadzi-Velkov, Slavche Pejoski, Jia Guo, Marija Poposka, and Arumugam Nallanathan


\thanks{Z. Hadzi-Velkov, S. Pejoski and M. Poposka are with the Faculty of Electrical Engineering and Information Technologies, Ss. Cyril and Methodius University, Skopje, North Macedonia (e-mail: \{zoranhv, slavchep, poposkam\}@feit.ukim.edu.mk).}  
\thanks{J. Guo is with the School of Electronic Engineering and Computer Science, Queen Mary University of London, UK (e-mail: jia.guo@qmul.ac.uk).} 
\thanks{A. Nallanathan is with the School of Electronic Engineering and Computer Science, Queen Mary University of London, UK and also with the Department of Electronic Engineering, Kyung Hee University, Yongin-si, Gyeonggi-do 17104, Korea (email: a.nallanathan@qmul.ac.uk).}
}

\markboth{ }{Shell \MakeLowercase{\textit{et al.}}: Bare IEEEtran.cls for Journals}
\maketitle

\begin{abstract}
Pinching-antenna systems (PASS) enable near-field beam focusing using radiating points along a dielectric waveguide, but coherent spherical-wave superposition makes their channels sensitive to small user displacements. This sensitivity is critical for power-domain non-orthogonal multiple access (NOMA), where successive interference cancellation (SIC) requires the weak-user layer to remain decodable at the strong user. We develop spatially robust PASS-NOMA designs that optimize antenna placement and power splitting over bounded service areas (SAs), with a preset decoding order and explicit SIC-feasibility constraints. We propose a worst-case design for quasi-static users and an ergodic design for codewords spanning representative SA positions, and characterize their achievable rate regions. Numerical results show that, despite SIC requiring weak-user decoding in both SAs, NOMA outperforms orthogonal multiple access (OMA) with a shared placement under both robustness models except when the strong user is served at low rate. In the ergodic regime, this advantage can persist even against OMA with per-user antenna reconfiguration. 
\end{abstract}


\vspace{-2mm}

\section{Introduction}
Pinching-antenna systems (PASS) are a flexible-antenna architecture in which small dielectric particles activated along a dielectric waveguide radiate the guided signal at selected locations~\cite{ding2025}. Since the waveguide aperture can span several meters, users are often located in the radiating near field, where the effective channel is governed by coherent spherical-wave superposition rather than by a single far-field steering direction~\cite{liu2025survey}. This property makes PASS attractive for power-domain non-orthogonal multiple access (NOMA), where near-field focusing can create the channel-gain disparity required by superposition coding and successive interference cancellation (SIC) that motivates a growing body of PASS-NOMA designs~\cite{xu2025,wang2025activation,xu2025qos,ren2026,gan2026}.

These studies mainly optimize rates, power, activation, or pinching locations assuming static users at known coordinates. However, near-field focusing creates a reliability issue that this fixed-coordinate design does not capture. In conventional far-field NOMA, the SIC order follows the users' large-scale channel gains and is stable under small displacements, because path loss changes slowly with position. In a PASS link, by contrast, the interference pattern can change substantially over a wavelength-scale displacement: a placement that makes the weak-user layer decodable at the strong user at the nominal position may, after a small displacement, leave the selected weak-user code rate above the cross-decoding rate available at the strong user. SIC fails in this case because the strong user cannot decode and cancel the weak-user layer. This fragility motivates a service area (SA) formulation of the PASS-NOMA design problem in which SIC feasibility is enforced over bounded user regions rather than only at nominal coordinates.

In this paper, we replace fixed user coordinates with bounded SAs to enable robust near-field NOMA communications. In the fixed-coordinate setting, the PASS capacity region is nearly triangular~\cite{ren2026,ouyang2026capacity}. Robust PASS designs under imperfect information have mainly been studied from a link-level beamforming perspective~\cite{sun2026}, while a recent geometry-aware network-level design uses a known blockage layout to configure one antenna per waveguide for region-wide single-user SNR coverage and worst-location performance~\cite{xu2026netII}. SA-based spatial robustness was recently considered for SWIPT in~\cite{hadzi2026swipt}. Unlike the single-user region-level SNR objectives in~\cite{xu2026netII} and the SWIPT formulation in~\cite{hadzi2026swipt}, our SA-based NOMA formulation must guarantee not only direct decoding within each user's SA but also decodability of the weak-user layer across the strong-user SA for SIC.

The contributions of this paper are twofold. ($1$) We formulate fixed-order near-field NOMA over SAs with explicit robust SIC-feasibility constraints, and develop quasi-static and ergodic designs, showing that the worst-case design splits power against the minimum channel gain over the SA while the ergodic design splits against its geometric mean. ($2$) We characterize the achievable rate regions of both designs, including the containment of the worst-case region within the ergodic one, and show that the NOMA gain over shared-placement OMA survives worst-case robustness except when the strong user is served at low rate, while ergodic averaging retains it over wider ranges of QoS targets and SA sizes.

\vspace{-1mm}

\section{System Model }
We consider downlink near-field NOMA with two users $U_1$ and $U_2$, in which a dielectric waveguide of length $L_w$ is deployed at height $h$ above the user plane $z=0$ along the $x$-axis and fed at its left end $q_0=-L_w/2$. A total of $N$ pinching antennas are activated at the positions $(q_n,0,h)$ along the waveguide, subject to a minimum inter-antenna spacing $\Delta$ to avoid electromagnetic coupling. If we collect $q_n$ in the vector $\mathbf q=[q_1,\ldots,q_N]^T$, the feasible placements form the set
\begin{equation}\label{eq:Qset}
\mathcal Q=\bigl\{\mathbf q\in[-L_w/2,L_w/2]^N:|q_n-q_{n'}|\ge\Delta,\forall n\neq n'\bigr\}.
\end{equation} 
The total radiated power is $P_T$, distributed equally among the $N$ active elements.

\emph{Service-area channel model:} Each user $U_k$ is associated with a bounded SA $\Hcal_k$ in the user plane, centered at $\mathbf c_k$, modeling users' positioning errors and mobility. The channel from $n$th pinching antenna at position $q_n$ ($1\leq n \leq N$) to $k$th user possible position coordinates $\mathbf r_k\in\Hcal_k$ ($k = 1, 2$) is 
\begin{equation} \label{channel}
 h_n(\mathbf r_k)=\frac{\eta}{d_n(\mathbf r_k)}
 \exp\!\left[-j\left(\frac{2\pi}{\lambda}d_n(\mathbf r_k)+\theta_n\right)\right],
\end{equation}
where $\eta=\lambda/(4\pi)$, $d_n(\mathbf r_k)$ is the Euclidean distance from the $n$th antenna to $\mathbf r_k$, and $\theta_n=(2\pi n_{\rm eff}/\lambda)(q_n-q_0)$ is the in-waveguide phase shift with $q_0$ the feed-point coordinate. Since the total power is equally divided among the active antennas, the received power gain is given by 
\begin{equation}
 g_k(\mathbf r_k,\mathbf q)=\frac{P_T}{N}\left|\sum_{n=1}^{N}h_n(\mathbf r_k)\right|^2. 
\end{equation}

We assume a fixed downlink decoding order $U_2\to U_1$, where $U_1$ is the nominally strong/SIC user, $U_2$ is the nominally weak user, such that $U_1$ decodes $U_2$'s layer before its own. The superposed baseband signal is
\begin{equation}
 s=\sqrt{\alpha}s_1+\sqrt{1-\alpha}s_2,\quad 0\le\alpha\le 1,
\end{equation}
where $\alpha$ is the power splitting ratio, i.e., the fraction of transmit power assigned to $U_1$. Given $\mathbf r_k\in\Hcal_k$, the weak-user rate, the cross-decoding rate, and the strong-user rate are  
\begin{align}
R_2(\mathbf r_2)&=\log_2\!\left(1+\frac{(1-\alpha)g_2(\mathbf r_2,\mathbf q)}{\alpha g_2(\mathbf r_2,\mathbf q)+\sigma^2}\right),\label{eq:R2}\\
R_{2\to1}(\mathbf r_1)&=\log_2\!\left(1+\frac{(1-\alpha)g_1(\mathbf r_1,\mathbf q)}{\alpha g_1(\mathbf r_1,\mathbf q)+\sigma^2}\right),\label{eq:R21}\\
R_1(\mathbf r_1)&=\log_2\!\left(1+\frac{\alpha g_1(\mathbf r_1,\mathbf q)}{\sigma^2}\right).\label{eq:R1}
\end{align}

\emph{CSI assumption}: The near-field LoS channel in~(\ref{channel}) is a deterministic function of the antenna placement $\mathbf q$ and the user position $\mathbf r_k$. At the design stage, the transmitter can evaluate the effective gains $g_k(\mathbf r_k,\mathbf q)$ on the SA grids. During actual operation, however, the instantaneous user position inside the SA is unavailable at the transmitter, since tracking wavelength-scale displacements of mobile users is impractical. The decoding order, antenna placement, power split, and scheduled rates are thus fixed by the design. Each receiver is assumed to have the local CSI required for coherent decoding and SIC. 

\emph{SIC feasibility:} In conventional NOMA, the decoding order is determined by the users' channel gains: the user with the weaker channel is decoded first and treated as the strong user's interference. Adaptive SIC ordering would require the transmitter to track both user positions at the wavelength scale. In the near field, however, these gains vary across each SA, so an ordering that holds at the nominal center $\mathbf{c}_k$ may be reversed at other positions in $\Hcal_k$. The decoding order is therefore preselected: both users decode the $U_2$ signal first, and $U_1$ cancels it before decoding its own signal, with appropriate constraints enforcing the validity of this order across the SAs.

\section{Robust Design Optimization} 
\label{sec:designs}


In the considered fixed-order NOMA system, the weak-user layer must be decodable by both users in the two SAs. Depending on the relation between the codeword duration and the user-mobility time scale, the same SA geometry leads either to a pointwise worst-case guarantee or to an ergodic rate criterion. In the first case (i.e., a \emph{quasi-static regime}), each codeword is decoded at a single but unknown position $\mathbf r_k\in\Hcal_k$, so reliable communication requires that all decoding events succeed at \emph{every} position in the service area, leading to worst-case rate guarantees. In an \emph{ergodic regime}, sufficiently long codewords, mobility, or interleaving expose the codeword to a representative sample of positions in $\Hcal_k$ ($k = 1,2$), so the relevant metric is the average mutual information over the SA. The spatial averages $\E_{\Hcal_k}[\cdot]$ in P2 are taken over a uniform position distribution on $\Hcal_k$, which is the maximum-entropy choice when only the SA is known. A nonuniform distribution can be also incorporated by weighting the grid points of $\mathcal S_k$, but it does not change the design, its analytical properties or the solution method. We design one scheme for each regime, P1 (worst-case NOMA) and P2 (ergodic NOMA), and compare them with five baselines, the pointwise NOMA (B1) and the OMA schemes (B2--B5). 

\vspace{-4mm}

\subsection{Worst-case NOMA scheme (P1)}

Quasi-static design maximizes guaranteed weak-user rate subject to strong-user QoS and SIC feasibility at each position: \vspace{-5mm}
\begin{subequations}\label{eq:P1}
\begin{align}
\max_{\mathbf q,\alpha,r_0}\quad & r_0\\
\text{s.t.}\quad
& \min_{\mathbf r_1\in\Hcal_1}R_1(\mathbf r_1)\ge R_1^{\min},\label{eq:P1b}\\
& \min_{\mathbf r_2\in\Hcal_2}R_2(\mathbf r_2)\ge r_0,\label{eq:P1c}\\
& \min_{\mathbf r_1\in\Hcal_1}R_{2\to1}(\mathbf r_1)\ge r_0,\label{eq:P1d}\\
& \mathbf q\in\mathcal Q,\quad 0\le\alpha\le1,\label{eq:P1e} 
\end{align}
\end{subequations}
where $\mathcal Q$ is defined by~\eqref{eq:Qset}. Constraint~\eqref{eq:P1b} guarantees the strong-user QoS throughout $\Hcal_1$, \eqref{eq:P1c} guarantees that $U_2$ decodes its own layer throughout $\Hcal_2$, and \eqref{eq:P1d} is the robust SIC constraint: the $U_2$ layer must remain decodable at $U_1$ across all of $\Hcal_1$.

\emph{Power splitting ratio.} For any fixed placement $\mathbf q$, $R_1$ in~\eqref{eq:R1} is increasing in $\alpha$, while $R_2$ and $R_{2\to1}$ in~\eqref{eq:R2}--\eqref{eq:R21} are decreasing. Since the $\min$-operators preserve this monotonicity, the weak-user rate $r_0$ is maximized by choosing the smallest $\alpha$ that satisfies~\eqref{eq:P1b} with equality. Evaluating~\eqref{eq:P1b} at the worst-case gain $g_1^{\min}(\mathbf q)=\min_{\mathbf r_1\in\Hcal_1}g_1(\mathbf r_1,\mathbf q)$ yields the closed form
\begin{equation}\label{eq:alpha_P1}
\alpha^{\rm P1}(\mathbf q)=\frac{\gamma_1\sigma^2}{g_1^{\min}(\mathbf q)},
\end{equation}
where $\gamma_1=2^{R_1^{\min}}-1$, which is thus an explicit function of the antenna placement. P1 is feasible iff $\alpha^{\rm P1}(\mathbf q)\le 1$, i.e., $g_1^{\min}(\mathbf q)\ge\gamma_1\sigma^2$. 

\vspace{-4mm}

\subsection{Ergodic NOMA scheme (P2)} 

In the ergodic regime the constraints hold in spatial average rather than at every position. P2 models delay-tolerant coded transmission to moving terminals, where the observations used for one decoding decision are collected over an interval long enough for both users to sample representative positions in their respective SAs. The ergodic design solves
\begin{subequations}\label{eq:P2}
\begin{align}
\max_{\mathbf q,\alpha,r_{\rm erg}}\quad & r_{\rm erg}\\
\text{s.t.}\quad
& \E_{\mathbf r_1\in\Hcal_1}[R_1(\mathbf r_1)]\ge R_1^{\min},\label{eq:P2b}\\
& \E_{\mathbf r_2\in\Hcal_2}[R_2(\mathbf r_2)]\ge r_{\rm erg},\label{eq:P2c}\\
& \E_{\mathbf r_1\in\Hcal_1}[R_{2\to1}(\mathbf r_1)]\ge r_{\rm erg},\label{eq:P2d}\\
& \mathbf q\in\mathcal Q,\quad 0\le\alpha\le1.\label{eq:P2e}
\end{align}
\end{subequations}
The averaged constraints in~\eqref{eq:P2b}--\eqref{eq:P2d} then follow from the standard ergodic-achievability argument: each decoder accumulates mutual information across the visited positions, and reliable decoding requires the code rate not to exceed the corresponding spatial average. In particular, unlike the per-position SIC requirement of P1,~\eqref{eq:P2d} ensures that $U_1$ accumulates sufficient mutual information to decode and cancel the $U_2$ layer even if the instantaneous gain ordering reverses at some positions, as in ergodic fading broadcast channels~\cite{li2001}.

\emph{Power splitting ratio.} For any feasible placement $\mathbf q$, the left-hand side of~\eqref{eq:P2b} increases with $\alpha$. Hence, the constraint is active at the optimum and uniquely determines $\alpha^{\rm P2}(\mathbf q)$. At high SNR, $R_1^{\min} \approx (1/L_1) \sum_{i=1}^{L_1} \log_2 (\alpha^{\rm P2}g_1(\mathbf r_i,\mathbf q)/\sigma^2) = \log_2 (\alpha^{\rm P2}\tilde g_1(\mathbf q)/\sigma^2)$, where $\tilde g_1(\mathbf q)=\bigl(\prod_{i=1}^{L_1}g_1(\mathbf r_i,\mathbf q)\bigr)^{1/L_1}$ is the geometric mean of the sampled channel gains. Therefore,
\begin{equation}\label{eq:alpha_P2_hsnr}
\alpha^{\rm P2}(\mathbf q) \approx
\frac{\gamma_1\sigma^2}{\tilde g_1(\mathbf q)},
\end{equation}
where $2^{R_1^{\min}} = 1 + \gamma_1 \approx \gamma_1$ at high SNR. Comparing~\eqref{eq:alpha_P2_hsnr} with~\eqref{eq:alpha_P1} gives $\alpha^{\rm P1}/\alpha^{\rm P2}\approx\tilde g_1(\mathbf q)/g_1^{\min}(\mathbf q)$, which quantifies the additional power available to $U_2$ through spatial averaging.

\begin{lemma}[Power-splitting ordering]\label{lem:alpha}
For any antenna placement $\mathbf q$, feasible for both P1 and P2, it holds that 
\begin{equation}\label{eq:alpha_order}
\alpha^{\rm P2}(\mathbf q)\le\alpha^{\rm P1}(\mathbf q).
\end{equation}
\end{lemma}

\begin{IEEEproof}
Per~\eqref{eq:alpha_P1}, the P1 satisfies $\min_{\mathbf r_1}R_1(\mathbf r_1)=R_1^{\min}$ at $\alpha=\alpha^{\rm P1}$, so $R_1(\mathbf r_1)\ge R_1^{\min}$ for every $\mathbf r_1\in\Hcal_1$. Averaging preserves the inequality, $\E_{\Hcal_1}[R_1]\ge R_1^{\min}$, so $\alpha^{\rm P1}$ is feasible for  average QoS constraint~\eqref{eq:P2b}. Since $\alpha^{\rm P2}$ is the smallest $\alpha$ satisfying~\eqref{eq:P2b} and $R_1$ increases in $\alpha$, it follows $\alpha^{\rm P2}\le\alpha^{\rm P1}$.
\end{IEEEproof}

\emph{Remark 1 (Rate-region nesting):} Since every spatial average dominates the corresponding worst case, any P1-feasible rate pair is also P2-feasible for the same antenna placement and power splitting ratio. Hence $\mathcal R_{\rm P1}\subseteq\mathcal R_{\rm P2}$, because P2's surplus power $(\alpha^{\rm P1} - \alpha^{\rm P2}) P_T$ (Lemma~\ref{lem:alpha}) is redirected to $U_2$.

\vspace{-2mm}

\subsection{Baseline Schemes}\label{subsec:baselines}

\emph{Pointwise NOMA scheme (B1)}: The baseline B1 scheme represents the conventional fixed-coordinate design that solves~\eqref{eq:P1} optimized for the nominal SA centers $\mathbf c_1$ and $\mathbf c_2$. The SIC order is fixed to $U_2 \to U_1$ based on the nominal channel ordering $g_1(\mathbf r_1,\mathbf q) >  g_2(\mathbf r_2,\mathbf q)$. B1 is optimistic relative to the service-area designs when $\mathbf c_1$ and $\mathbf c_2$ coincide with near-field focusing spots. For arbitrary geometries it is a benchmark rather than a bound, since the center gain need not dominate the worst-case or average gain. To expose its fragility, we also average the rates optimized for the fixed locations $\mathbf c_1$ and $\mathbf c_2$ over the full service areas, $\bar R_k=\E_{\Hcal_k}[R_k]$ ($k = 1, 2$), which measures the average rate that can be delivered from each SA. When averaging $\bar R_1$, if $R_{2\to1}(\mathbf r_1)$ falls below the weak-user rate at the nominal centers, $U_1$ cannot cancel the $U_2$ layer, and its pointwise rate is computed by treating that layer as residual interference.

\emph{Worst-case OMA scheme (B2)}: B2 can help identify the gain of NOMA superposition over orthogonal access. It avoids SIC constraints but sacrifices spectral efficiency due to orthogonal time sharing, with a single antenna placement $\mathbf q$ shared by both time phases, matching the shared-placement assumption of P1 and P2. Let $\tau$ and $1-\tau$ denote the fraction of time allocated to $U_1$ and $U_2$, respectively, where $0 \leq \tau \leq 1$. The single-user capacities are calculated as worst-case rates,
\begin{equation}\label{eq:Ck}
C_k(\mathbf q) \triangleq \min_{\mathbf r_k\in\Hcal_k}\log_2\!\left(1+g_k(\mathbf r_k,\mathbf q)/\sigma^2\right), \quad k = 1, 2,
\end{equation}
and the placement is optimized to maximize the weak-user rate $(1-\tau) C_2(\mathbf q)$ subject to $\tau C_1(\mathbf q) \geq R_1^{\min}$.

\emph{Ergodic OMA scheme (B3)}: B3 is the OMA counterpart of P2 under the same shared placement. The single-user capacities are the spatial averages $\bar C_k(\mathbf q)\triangleq\E_{\mathbf r_k\in\Hcal_k}[\log_2(1+g_k(\mathbf r_k,\mathbf q)/\sigma^2)]$, $k=1,2$, and the placement maximizes $(1-\tau)\bar C_2(\mathbf q)$ subject to $\tau\bar C_1(\mathbf q)\ge R_1^{\min}$. 

\emph{Reconfigurable OMA schemes (B4, B5):} Let us define the optimized worst-case, spatial-average, and center-point single-user capacities ($k=1, 2$), 
\begin{align}
C_k^{\rm w}&\triangleq\max_{\mathbf q\in\mathcal Q}\;C_k(\mathbf q),\label{eq:Ckmax}\\
C_k^{\rm erg}&\triangleq\max_{\mathbf q\in\mathcal Q}\;\E_{\mathbf r_k\in\Hcal_k}\!\left[\log_2\!\left(1+\frac{g_k(\mathbf r_k,\mathbf q)}{\sigma^2}\right)\right],\label{eq:Ckerg}\\
C_k^{\rm c}&\triangleq\max_{\mathbf q\in\mathcal Q}\;\log_2\!\left(1+\frac{g_k(\mathbf c_k,\mathbf q)}{\sigma^2}\right),\label{eq:Ckc}
\end{align}
where $C_k(\mathbf q)$ is defined by~\eqref{eq:Ck}. These satisfy $C_k^{\rm w}\le C_k^{\rm erg}$ and $C_k^{\rm w}\le C_k^{\rm c}$, with equality as $\delta\to0$.

If the placement can be re-optimized for each time phase, the two phases decouple and each uses its own single-user-optimal placement. The rate regions are then the time-sharing segments between $(C_1^{\rm w},0)$ and $(0,C_2^{\rm w})$ under the worst-case criterion (B4), and between $(C_1^{\rm erg},0)$ and $(0,C_2^{\rm erg})$ under the ergodic criterion (B5). B4 and B5 upper-bound B2 and B3, respectively, but require mechanical repositioning of the PAs between the phases. 

\vspace{-2mm}

\subsection{Boundaries of the Achievable Rate Regions} \label{subsec:corners}
The rate-region boundary of each scheme connects a left endpoint (largest weak-user rate at $R_1^{\min}=0$) with a right endpoint (largest supportable strong-user target as the weak-user rate vanishes). 

\emph{Right endpoints ($R_2\to0$):} Each scheme reduces to maximizing the strong-user rate under its own spatial metric: P1, B2, and B4 terminate at $(C_1^{\rm w},0)$, P2, B3, and B5 at $(C_1^{\rm erg},0)$, and B1 at $(C_1^{\rm c},0)$. In particular, P1 and B2 share the same right endpoint.

\emph{Left endpoints ($R_1^{\min}=0$):} For the OMA schemes and B1, only the weak user matters: B2 and B4 reach $(0,C_2^{\rm w})$, B3 and B5 reach $(0,C_2^{\rm erg})$, and B1 reaches $(0,C_2^{\rm c})$. For P1, $\alpha\to0$ but the SIC constraint~\eqref{eq:P1d} remains active, so the weak-user layer must be decodable over both $\Hcal_1$ and $\Hcal_2$ simultaneously:
\begin{equation}\label{eq:P1_left}
\begin{aligned}
R_2^{{\rm w},\rm P1}\triangleq\max_{\mathbf q\in\mathcal Q}\,\min\Bigl\{&\min_{\mathbf r_2\in\Hcal_2}\!\log_2\!\left(1+\frac{g_2(\mathbf r_2,\mathbf q)}{\sigma^2}\right),\\
&\min_{\mathbf r_1\in\Hcal_1}\!\log_2\!\left(1+\frac{g_1(\mathbf r_1,\mathbf q)}{\sigma^2}\right)\Bigr\},
\end{aligned}
\end{equation}
which satisfies $R_2^{{\rm w},\rm P1}\le C_2^{\rm w}$: covering two areas with one placement cannot beat covering one. The P2 left endpoint $R_2^{{\rm e},\rm P2}$ is defined analogously with expectations in place of the minima, and satisfies $R_2^{{\rm w},\rm P1}\le R_2^{{\rm e},\rm P2}\le\min\{C_1^{\rm erg},C_2^{\rm erg}\}$ by Remark~1. It may exceed $C_2^{\rm w}$ because averages are not pointwise guarantees.

\vspace{-2mm}

\subsection{Solution Method}\label{subsec:solution}

The PA positions are modeled over the continuous set $\mathcal Q$, and the solver restricts them to the discrete subset $\mathcal Q_{\Delta_c}\subseteq\mathcal Q$, where the candidate spacing $\Delta_c$ is a numerical resolution parameter distinct from the physical separation $\Delta$. The waveguide is divided into $M = \lfloor L_w/\Delta_c\rfloor+1$ equidistant candidate antenna locations with spacing $\Delta_c$, and the placement is represented by a binary activation vector $\mathbf a\in\{0,1\}^M$ with the cardinality condition $\mathbf 1^T\mathbf a=N$. The candidate spacing is set to $\Delta_c=\Delta=\lambda/2$, matching the minimum inter-antenna spacing. Each SA $\Hcal_k$ is discretized into a grid $\mathcal S_k$ of $L_k$ points at spacing $\Delta_g$ ($k=1,2$). The worst-case operators become minima over $\mathcal S_k$, expectations become sample averages, and the reported worst-case guarantees are grid-based guarantees over $\mathcal S_k$. The channel matrices $\mathbf H_k\in\mathbb C^{L_k\times M}$ from all antenna candidate positions to all grid points are precomputed once, and the design reduces to selecting $N$ of the $M$ candidates, which leads to a combinatorial selection over $\binom{M}{N}$ feasible subsets\footnote{A grid-resolution study, in which the P1 and P2 designs were re-evaluated on a $\lambda/8$ reference grid and by a continuous search for spatial nulls within each SA, shows that the $\lambda/2$ design grid overestimates the worst-case rates by less than 1\% along the rate-region boundary, with P2 being less sensitive than P1.}.

The antenna placement is carried out by one-swap local search motivated by~\cite{wang2025activation,hadzi2026swipt}. Each pass examines all $N(M-N)$ placements obtained by exchanging one active PA for one unused candidate location, accepts improving exchanges, and stops when no exchange improves the objective. As an exchange changes only one PA, the effective channel is updated incrementally at $O(L_k)$ cost, and one pass costs $O(N(M-N)(L_1+L_2))$. The search is reinforced by multi-start from a number of fixed placements (gain-ranked, window, and pseudorandom seeds from a fixed generator state), warm starts from the single-user solutions and from the neighboring values of $R_1^{\min}$, and re-optimization after perturbing exchanges. For P1, each placement is evaluated with the closed-form $\alpha^{\rm P1}$. For P2, $\alpha^{\rm P2}$ is computed by Newton iterations initialized at $\alpha^{\rm P2}$ given by~\eqref{eq:alpha_P2_hsnr}.


\section{Numerical Results}
\label{sec:results}

For the numerical analysis, unless otherwise stated, we set \(f_c=28\)~GHz, \(n_{\rm eff}=1.4\), \(L_w=6\)~m, \(h=1.5\)~m, \(N=40\), \(P_T=10\)~W, and \(\sigma^2=10^{-7}\)~W. The minimum inter-antenna spacing, PA-candidate spacing, and SA-grid spacing are all set to \(\lambda/2\), i.e., \(\Delta=\Delta_c=\Delta_g=\lambda/2\). Both SAs are disks of radius \(\delta=10\)~cm, centered at \(\mathbf c_1=(-1,0.8,0)\)~m and \(\mathbf c_2=(1.5,-2,0)\)~m, respectively. Thus, \(U_1\) is closer to the waveguide than \(U_2\), providing the nominal channel-gain imbalance underlying the prescribed NOMA decoding order.

\begin{figure}[t]\centering
\includegraphics[width=0.97\columnwidth, height=0.25\textheight, keepaspectratio=false]{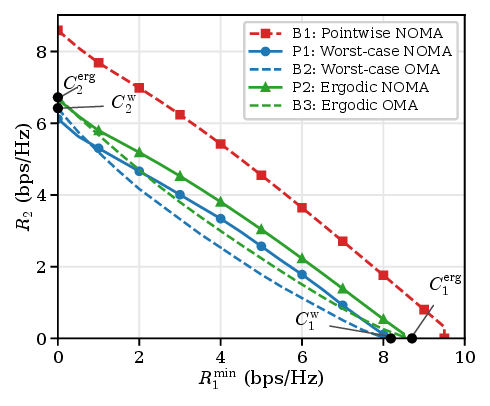} 
\vspace{-5mm}
\caption{Rate region (NOMA vs. OMA). \vspace{-8mm}}
\label{fig:rr}
\end{figure}

\begin{figure}[t]\centering
\includegraphics[width=0.97\columnwidth, height=0.25\textheight,keepaspectratio=false]{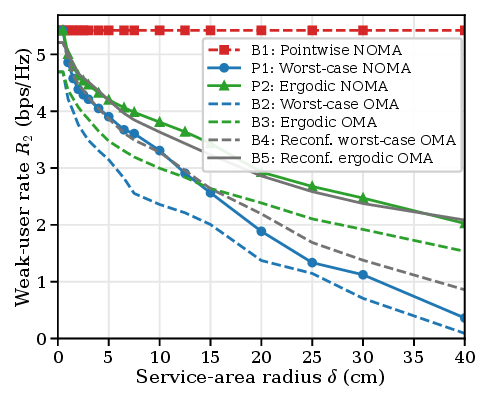} \vspace{-5mm}
\caption{Weak-user rate vs.\ service-area radius $\delta$ at $R_1^{\min}=4$~bps/Hz. \vspace{-5mm}} 
\label{fig:delta}
\end{figure}

\emph{Rate region (Fig.~\ref{fig:rr}).} The pointwise design B1 lies above the robust schemes in this geometry because it optimizes only the SA centers; the gap between the B1 and P1/P2 boundaries is the rate penalty to guarantee performance over an SA rather than at a single point. Under the worst-case criterion, P1 and B2 share the right endpoint $(C_1^{\rm w},0)$, but B2 leads at low $R_1^{\min}$: the SIC constraint~\eqref{eq:P1d} forces a single placement to cover both $\Hcal_1$ and $\Hcal_2$ even though $U_1$ carries little rate, whereas OMA needs only single-area coverage per time slot, so the P1 left endpoint falls short of $C_2^{\rm w}$ by the gap $C_2^{\rm w}-R_2^{{\rm w},\rm P1}\ge0$. Near the shared endpoint the ordering reverses. Both schemes are then confined to the placement $\mathbf q^\star$ achieving $C_1^{\rm w}$, where B2's weak-user rate $(1-R_1^{\min}/C_1^{\rm w})\,C_2(\mathbf q^\star)$ vanishes linearly because time sharing surrenders a growing fraction of time to $U_1$, whereas P1 serves $U_1$ in the power domain and retains the full time resource for $U_2$. Whenever $g_2^{\min}(\mathbf q^\star)<g_1^{\min}(\mathbf q^\star)$, P1's rate decays more slowly than B2's, so P1 lies above B2 there and the two boundaries cross by continuity. Under the ergodic criterion, P2 exceeds B3 throughout the interior of the region. The near-linearity of the boundaries is consistent with the high-SNR behavior of a fixed, ordered two-user downlink: both the NOMA and OMA boundaries approach slope $-1$, while their separation is governed by the channel-gain disparity and becomes small relative to the overall rates, consistent with the near-triangular PASS capacity regions for fixed user coordinates in~\cite{ouyang2026capacity}.

\begin{figure}[t]\centering
\includegraphics[width=0.9\columnwidth]{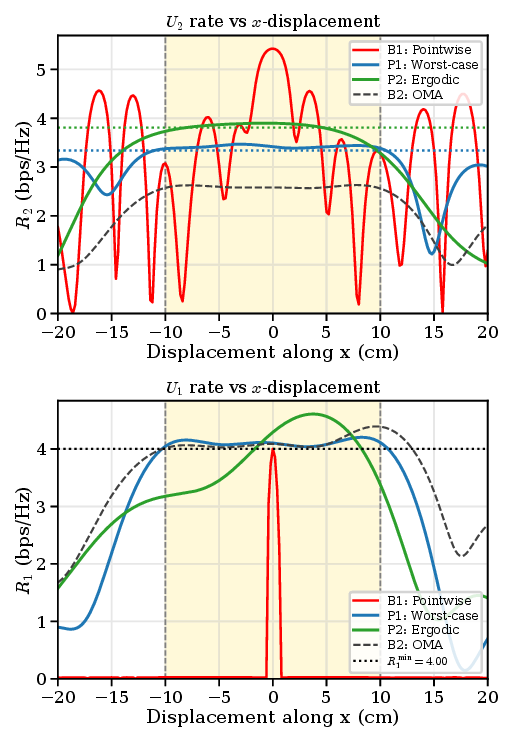} \vspace{-3mm}
\caption{Spatial resilience at $R_1^{\min}=4$~bps/Hz: achievable rates vs.\ user displacement from the SA center along $x$ axis (shaded band marks the SA). \vspace{-5mm} }
\label{fig:res}
\end{figure}

\emph{Service-area size (Fig.~\ref{fig:delta}).} This figure illustrates how the robustness model reacts to increasing spatial uncertainty. As $\delta\to0$ the SAs collapse to points, P1 and P2 converge to the pointwise NOMA design, and the classical NOMA advantage over OMA reappears. As $\delta$ grows, P1 and B2 decay steeply and remain close because both depend on the worst-case gain, which degrades rapidly as larger SAs admit deeper near-field nulls. P1 stays above B2 for all $\delta$ and above the reconfigurable B4 up to $\delta\approx13$~cm. P2 and B3 decay more gracefully because a spatial average is insensitive to isolated nulls. P2 remains at or above the reconfigurable B5 over the entire range, with a visible margin beyond $\delta\approx10$~cm. The ergodic criterion thus preserves the NOMA advantage over reconfigurable OMA up to about three times larger SA radii than the worst-case criterion. The P2--P1 gap widens with $\delta$, reflecting the growing benefit of spatial averaging over larger SAs.

\emph{Spatial resilience (Fig.~\ref{fig:res}).} This figure shows how sensitive each design is to user displacement. B1 delivers its rates only at the exact center. Its $R_2$ follows the near-field interference pattern, oscillating between peaks and deep nulls. Its $R_1$ is a narrow peak at $\mathbf c_1$: away from the center, SIC fails and the weak-user signal remains as strong interference, so $R_1$ drops to near zero. P1 stays nearly flat at or above its guarantee $r_0^\star$ throughout the SA, because maximizing the worst case equalizes the rate across the region. P2 achieves a higher rate, but only on average: its profile drops below the average near the SA edges, which the ergodic constraints permit. B2 is also nearly flat inside the SA, at a lower level due to the time-sharing cost. In summary, P1 guarantees a minimum rate at every position, whereas P2 guarantees a higher averaged rate.

\vspace{-1mm}

\section{Conclusion}
We studied spatially robust near-field NOMA with pinching antennas, in which the antenna placement and power splitting are optimized to serve bounded service areas rather than fixed user coordinates. Under a fixed decoding order, SIC feasibility couples the two SAs through one placement. It costs the worst-case design its advantage over shared-placement OMA only when the strong user is served at low rate. Otherwise the superposition gain survives, though marginal against reconfigurable OMA. Ergodic averaging removes this limitation, preserving the gain over wider QoS targets and SA sizes even against reconfigurable OMA. Pointwise designs deliver their promised rates only near the exact nominal positions and should not be interpreted as SA guarantees.


\vspace{-2mm}

\begin{thebibliography}{99}

\bibitem{ding2025}
Z.~Ding, R.~Schober, and H.~V.~Poor,
``Flexible-antenna systems: A pinching-antenna perspective,''
\emph{IEEE Trans. Commun.}, vol.~73, no.~10, pp.~9236--9253, Oct.~2025.

\bibitem{liu2025survey}
Y.~Liu, C.~Ouyang, Z.~Wang, J.~Xu, X.~Mu, and A.~L.~Swindlehurst,
``Near-field communications: A comprehensive survey,''
\emph{IEEE Commun. Surveys Tuts.}, vol.~27, no.~3, pp.~1687--1728, 2025.

\bibitem{xu2025}
Y.~Xu, Z.~Ding, and G.~K.~Karagiannidis,
``Rate maximization for downlink pinching-antenna systems,''
\emph{IEEE Wireless Commun. Lett.}, vol.~14, no.~5, pp.~1431--1435, May~2025.

\bibitem{wang2025activation}
K.~Wang, Z.~Ding, and R.~Schober,
``Antenna activation for NOMA assisted pinching-antenna systems,''
\emph{IEEE Wireless Commun. Lett.}, vol.~14, no.~5, pp.~1526--1530, May~2025.

\bibitem{xu2025qos}
Y.~Xu, Z.~Ding, D.~Cai, and V.~W.~S.~Wong,
``QoS-aware NOMA design for downlink pinching-antenna systems,''
\emph{IEEE Trans. Commun.}, vol.~73, no.~12, pp.~13611--13625, Dec.~2025.

\bibitem{ren2026}
Q.~Ren, X.~Mu, S.~Lin, and Y.~Liu,
``Pinching-antenna systems (PASS) meet multiple access: NOMA or OMA?''
\emph{IEEE Trans. Veh. Technol.}, early access, 2026,
doi:~10.1109/TVT.2026.3683670.

\bibitem{gan2026}
D.~Gan, X.~Xu, J.~Zuo, X.~Ge, and Y.~Liu,
``Joint beamforming for NOMA assisted pinching antenna systems (PASS),''
\emph{IEEE Trans. Commun.}, vol.~74, pp.~2450--2465, 2026.

\bibitem{hadzi2026swipt}
Z.~Hadzi-Velkov, M.~Poposka, S.~Pejoski, and A.~Nallanathan, ``Spatially robust near-field SWIPT using pinching antennas: Rate-energy tradeoff bounds,'' \emph{IEEE Wireless Commun. Lett.}, early access, 2026, doi:~10.1109/LWC.2026.3700698.

\bibitem{sun2026}
M.~Sun, X.~Mu, C.~Ouyang, G.~Shang, and Y.~Liu,
``Robust beamforming for pinching-antenna systems-based multi-user
communications,''
\emph{IEEE Wireless Commun. Lett.}, vol.~15, pp.~2859--2863, 2026.


\bibitem{li2001}
L.~Li and A.~J. Goldsmith, ``Capacity and optimal resource allocation for
fading broadcast channels---Part I: Ergodic capacity,'' \emph{IEEE Trans.
Inf. Theory}, vol.~47, no.~3, pp. 1083--1102, Mar. 2001. 

\bibitem{ouyang2026capacity}
C.~Ouyang, Z.~Wang, Y.~Liu, H.~Shin, and Z.~Ding,
``Capacity characterization of pinching-antenna systems,''
\emph{IEEE Trans. Wireless Commun.}, vol.~25, pp.~10387--10404, 2026,
doi:~10.1109/TWC.2026.3653191.

\bibitem{xu2026netII} Y.~Xu, Z.~Ding, X.~Y.~Zhang, T.~Q.~Duong, and T.-H.~Chang, ``Environment-aware network-level design of generalized pinching-antenna systems---Part~II: Geometry-aware case,'' arXiv:2602.17032.


\end{thebibliography}
\end{document}